\input{psfig.sty}

\documentclass{emulateapj}
\usepackage[light,all]{draftcopy}
\usepackage{apjfonts}

\usepackage{graphicx}
\usepackage{natbib}
\usepackage{amsmath}           
\usepackage{amsfonts}           

\slugcomment{Accepted for publication in The Astrophysical Journal}
\shorttitle{Hydrodynamic model of foot-point heated loops}
\shortauthors{Testa et al.}

\begin{document}

\title{Emission measure distribution in loops impulsively
	heated at the foot-points}

\author{Paola Testa\altaffilmark{1}, Giovanni Peres\altaffilmark{1},
	Fabio Reale\altaffilmark{1}}
\altaffiltext{1}{Dipartimento di Scienze Fisiche ed Astronomiche, 
	Sezione di Astronomia, Universit\`a di Palermo, 
	Piazza del Parlamento 1, 90134 Palermo, Italy}

\begin{abstract}
This work is prompted by the evidence of sharply peaked emission 
measure distributions in active stars, and by the claims of isothermal 
loops in solar coronal observations, at variance with the predictions 
of hydrostatic loop models with constant cross-section and uniform 
heating.
We address the problem with loops heated at the foot-points. 
Since steady heating does not allow static loop models solutions, we 
explore whether pulse-heated loops can exist and appear as steady loops, 
on a time average.    
We simulate pulse-heated loops, using the Palermo-Harvard 1-D 
hydrodynamic code, for different initial conditions corresponding 
to typical coronal temperatures of stars ranging from intermediate 
to active ($T \sim 3$--$10 \times 10^6$~K).
We find long-lived quasi-steady solutions even for heating 
concentrated at the foot-points over a spatial region of the order 
of $\sim 1/5$ of the loop half length and broader.
These solutions yield an emission measure distribution with a peak
at high temperature, and the cool side of the peak is as steep 
as $\sim T^{5}$, in contrast to the usual $\sim T^{3/2}$ of  
hydrostatic models with constant cross-section and uniform heating.
Such peaks are similar to those found in the emission measure
distribution of active stars around $10^7$~K. 
\end{abstract}

\maketitle

\section{Introduction}
\label{intro}

Coronae are important in the study of solar and stellar physics 
for several reasons: they are good tracers of stellar activity and 
of dynamo phenomena; also, bright coronae identify young stars and 
stellar formation regions more easily than many other characteristics.
Nevertheless, fundamental aspects of stellar coronae---namely the 
heating mechanisms that sustain the hot, confined plasma---are not 
well understood.
Although we are not able to observe directly the heating processes
at work in stellar coronae, the characteristics of the observed 
coronal structures can provide us with an indirect probe for the 
properties of the coronal heating.

The earliest spatially resolved observations of the solar corona
\citep[e.g.][]{Vaiana73} have already showed that the hot plasma 
is highly structured and confined by the magnetic field in loop 
structures, which are considered the basic building blocks of the 
coronae. 
Considerable effort has been devoted in the last three decades to 
understanding the physics of these structures of confined plasma and 
to develop adequate models that account for the observed properties 
of coronal emission.
The first loop models (e.g., \citealp{RTV}, hereafter RTV;
\citealp{Vesecky79,S81}, hereafter S81) were flux tubes of constant
cross section filled with plasma in hydrostatic equilibrium, and in 
energy balance under the effects of steady heating, heat flux and 
radiative losses, with the magnetic field only confining the plasma.
These models demonstrated a wide range of validity and satisfactorily 
reproduced a large number of coronal X-ray and EUV observations; both 
solar and stellar \citep[e.g.][]{Rosner77,Pallavicini81,Giampapa85,
Landini85,Peres87,Reale88,Reale02loop,Testa02}. 
However {\em TRACE} and {\em SoHO} have observed loops apparently 
incompatible with hydrostatic equilibrium (in terms of their spatial 
distribution of temperature and density) though appearing as 
quasi-static \citep[e.g.][]{Brekke97,Warren02,Winebarger02,Golub04}.
Some authors claimed that these loops are heated non-uniformly and 
in particular at their foot-points 
\citep{Aschwanden00,Aschwanden01,Winebarger03}. On the other hand,
if the heating is too localized at the foot-points, static loops 
should be thermally unstable, as widely discussed since the first 
works on modeling of loop structures, e.g.\ RTV, \cite{Antiochos79}, 
S81, \cite{Peres82}. 
Recently, detailed analyses of solar observations have brought up
again the question of the location of the heating release
\citep[e.g.][]{Priest00,Asch01loop,Reale02loop}.

In order to attempt to explain structures that apparently persist
for time scales longer than the characteristic cooling time with 
foot-point heating models, the models must be dynamic, since static 
solutions with foot-point heating can be unstable.
Dynamic models of this kind have been recently used in several works 
\citep[e.g.][]{Warren02,Warren03,Spadaro03,Mueller04}.  For instance,
\cite{Warren03} successfully reproduced several observed
characteristics with a multi-threaded model, impulsively heating 
loops to several million degrees and allowing them to cool to 
{\em TRACE}--observable temperatures ($\sim 10^6$~K).

The study of the stellar coronae allows us to investigate the effect
of stellar parameters (effective temperature, surface gravity, 
rotation, chemical composition, etc) on the coronal models, developed 
for the Sun.
On one hand, when studying stellar coronae a first approach is based 
on the hypothesis that the solar corona is an adequate paradigm to 
interpret the observations of stellar coronae.  On the other hand, 
the assumption of the solar analogy must be validated by comparing 
the characteristics of solar and stellar coronae.

Recent high quality spectral observations have provided us with 
detailed information on the properties of stellar coronal emission. 
High resolution spectra obtained with, e.g., {\em EUVE}, in the EUV 
range, and with the new observatories {\em Chandra} and 
{\em XMM-Newton} in the X-ray band, allow us to diagnose the plasma 
conditions in a large sample of stellar coronae at different activity 
levels.  
Since stellar coronae cannot be spatially resolved by present-day
telescopes, we must resort to indirect means in order to compare the 
properties of coronal structures among different stars and with the
observed characteristics of the solar corona.
The {\em emission measure distribution vs. temperature}, $EM(T)$, 
defined by:
\begin{equation} \label{eq:EM}
EM(T_i) = \int_{\Delta T_i} n_e^2(T) dV,
\end{equation}
where $n_e$ is the electron density, and V is the volume filled with
plasma at $T \in \Delta T_i$, contains substantial information about 
the coronal emitting plasma, mainly on its thermal structuring, and 
proves to be a useful diagnostic tool for comparing the coronal 
emission of different stars.

The global $EM(T)$ of the X-ray Sun is typically peaked at 
$T \sim 2$--$3 \times 10^6$~K and the ascending cool part is 
characterized by a rise as $\sim T^{3/2}$ \citep{Orlando00,Peres00}, 
even though its specific properties can vary for different coronal 
regions and in different phases of activity. 
Previous studies of the solar atmosphere, mostly done in the UV band 
\citep{Jordan80,Brosius96,Landi97}, led to analogous results. 
The observed dependence on temperature is well explained in terms of 
hydrostatic loops \citep{Peres01}: for a standard RTV uniformly heated 
loop model, the $T$ and $n$ structuring of the plasma along the loop 
yields $EM(T)$ increasing approximately as $T^{3/2}$. Thus, for a 
corona of optically thin plasma, mostly confined in hydrostatic loops, 
this would also yield $EM(T) \propto T^{3/2}$ in the ascending 
part. The descending part of $EM(T)$ gives us information on the 
distribution of the hydrostatic loops. 
Therefore, in the stellar case, the analysis of the global $EM(T)$
can provide us with information on the structuring of the observed 
corona, and can allow us to test the hypothesis of a corona formed 
by static loops at different maximum temperature.

Analyses of $EM(T)$ derived from EUV and X-ray spectra of 
several stars have appeared in the recent literature (e.g., 
\citealp{Sanz02,Sanz03a}, present extensive $EM(T)$ reconstructions 
from {\em EUVE} spectra of active stars), and most of them show
similar features.  
The $EM(T)$ of active stars is typically characterized by an 
ascending part as $T^{\alpha}$ with $\alpha > 3/2$, and up 
to $\sim 5$ in some cases 
\citep[see e.g.][]{Dupree93,Griffiths98,Drake00,Mewe01,Sanz02,
Sanz03a,Argi03,Scelsi04}. Moreover, the $EM(T)$ of several active 
coronae shows prominent "bumps", i.e.\ large amounts of almost 
isothermal material \citep[e.g.][]{Drake96,Dupree96,Dupree02}.
For example, \cite{Dupree93} used {\em EUVE} data for the giant 
Capella and derived an $EM(T)$ with a well defined and narrow peak 
at $\log(T) \sim 6.8$, and a rise much steeper than $T^{3/2}$; 
the analysis of X-ray {\em Chandra} spectra of Capella 
\citep{Mewe01,Argi03} has confirmed these findings.
Although these results are quite controversial and are based on 
atomic physics parameters still under refinement, they point to 
an almost isothermal hot part of the corona which, if formed 
by loops, does not seem to be compatible with the predictions of 
simple static loop models with uniform cross-section.

Isothermal loops would have fundamental implications for the 
thermodynamics and for the heating of the magnetized plasma.
In uniformly heated loops, the energy radiated by the plasma near
to the foot-points (that is at $T \sim 10^5$) is balanced by the 
local heat deposition and by the heat flux from the hotter plasma 
further up the loop.
However, for foot-point heating, the local energy deposition can 
supply most of the energy radiated by the cooler plasma, so that a 
lower heat flux (shallower temperature gradient) is needed from the 
hotter plasma and the $T$ profile becomes flatter.
Since standard hydrostatic loop models do not allow stable solutions 
for heating deposition over a spatial region smaller than about 
$L/3$ (e.g.\ S81), we decided to explore the possibility of 
obtaining stable loops with a non-constant and thus dynamic heating.

The possibility of a corona composed of loops heated episodically,  
close to their foot-points, poses two basic questions:
\begin{itemize}
\item Under what conditions can such loops be stable for time scales
much longer than the characteristic cooling time?
\item If long-lived, foot-point heated loops exist, is their
$EM(T)$ different from that of uniformly heated loops?
\end{itemize}

In this work we model the properties of foot-point heated coronal
loops, using a 1-D hydrodynamic model.  We run simulations with 
pulse-heating concentrated at the foot-points, and investigate the
existence of long-lived solutions; we study the $EM(T)$ associated 
with these solutions. In particular, we investigate the existence 
of stable loops characterized by an $EM(T)$ with a narrow peak and
a slope steeper than $T^{3/2}$, i.e.\ characteristic of those derived 
from observations of active stars.  We note that, for the solar case, 
\citet[see also \citealp{CK97,KC01}]{CK04} demonstrated that the 
$EM(T)$ provides a useful diagnostic for periodically heated plasma, 
by using a zero-dimensional model (i.e., each loop is characterized 
by a single, averaged, value of density and temperature) that 
considers nanoflare heating \citep{Cargill94}.

In \S\ref{sec2} we describe the loop modeling and the properties 
of the simulations. In \S\ref{sec3} we analyze the results of the
simulations. In \S\ref{sec4} we discuss our results and their 
implications, and then we draw conclusions from our work.

\section{Loop Model and Set of Simulations}
\label{sec2}

We model the plasma confined in a single magnetic flux tube.
We will discuss later how this single loop model can be useful for 
interpreting the overall emission from a corona.

In our model, the magnetic field has only the role of confining the
plasma; thermal conduction is effective only along the magnetic field
lines.  The model assumes constant cross-section along the loop.  
In our study, the loop half length is assumed to be $L=10^{10}$~cm, 
as observed for relatively long loops on the Sun. We discuss below 
the implications of our assumptions.

We simulate coronal loops heated close to their foot-points by
episodic pulses. 
In order to investigate the effect of some basic parameters on the 
solutions for the loop plasma, we have reduced the number of free 
parameters as much as possible. For instance, we assume identical
conditions in both loop legs and, therefore  the loop model
is symmetric about the loop apex. Because of this symmetry, the 
equations for the loop plasma are solved for half of the loop only.

The simulations are long lasting to search for long-term stability, 
i.e.\ for loops settling in a state steady on average, if such 
solutions exist.

\begin{deluxetable*}{lrrrrrccccr}
\tablecolumns{10} 
\tabletypesize{\footnotesize}
\tablecaption{Parameters of the simulations. \label{tabmod}}
\tablewidth{0pt}
\tablehead{
 \colhead{model \tablenotemark{a}} & \colhead{} & 
 \multicolumn{4}{c}{Initial conditions} & \colhead{} & 
 \multicolumn{3}{c}{Heat pulses \tablenotemark{b}} & \colhead{} \\
 \cline{3-6} \cline{8-10} \\[-0.3cm]
 \colhead{} & \colhead{$L$ \tablenotemark{c}} & 
 \colhead{$T_{\rm max}$ \tablenotemark{d}} & 
 \colhead{$p_{\rm base}$ \tablenotemark{e}} & 
 \colhead{$E_0$ \tablenotemark{f}} &
 \colhead{$\tau_{\rm cool}$ \tablenotemark{g}} & \colhead{} & 
 \colhead{$\sigma$ \tablenotemark{h}} & 
 \colhead{$\langle E_H \rangle$ \tablenotemark{i}} & 
 \colhead{$\tau$ \tablenotemark{l}} & 
 \colhead{$t_{\rm run}$ \tablenotemark{m}} 
 }
\startdata
 C & 1 & 3 & 1 & 0.45 & $\sim 2200$ & & 
 	L/3, L/5, L/10  & $E_0$, 4$E_0$  & 
	$\tau_{\rm cool}$/4, $\tau_{\rm cool}$/2 & 10\,000 \\ 

 H & 1 & 10 & 36 & 30 & $\sim 1200$ & & 
 	L/3, L/5, L/10  & $E_0$, 4$E_0$  & 
	$\tau_{\rm cool}$/4, $\tau_{\rm cool}$/2 & 5\,000 \\ \vspace{-0.2cm}
\enddata 
\tablenotetext{a}{C and H indicate the models of cooler 
	($T = 3 \times 10^6$~K) and hotter loops ($T = 10^7$~K) 
	respectively.}
\tablenotetext{b}{The set of simulations considers all possible 
	combinations of the values of the parameters ($\sigma$, 
	$\langle E_H \rangle$, $\tau$).}
\tablenotetext{c}{ {Loop} semilength in units of $10^{10}$~cm.}
\tablenotetext{d}{Maximum temperature (= apex temperature) in units 
	of $10^6$~K.}
\tablenotetext{e}{Base pressure in units of dyn/cm$^2$.}
\tablenotetext{f}{Heating per unit time and per unit volume 
	(erg~cm$^{-3}$~s$^{-1}$) from RTV loop scaling laws.}
\tablenotetext{g}{ {Loop} cooling time (see Eq.~\ref{eq:tau}) in seconds.}
\tablenotetext{h}{Gaussian parameter of the spatial extent 
		of the heating (see Eq.~\ref{eq:gauss}).}
\tablenotetext{i}{Heating intensity averaged in time and along the loop, 
		expressed in terms of the static heating, $E_0$.}
\tablenotetext{l}{Period of the heat pulses.}
\tablenotetext{m}{Total simulation time (s).}
\end{deluxetable*}

We use the Palermo-Harvard code \citep{Peres82,Betta97}; a 1-D
hydrodynamic code that consistently solves the time-dependent
density, momentum and energy equations for the plasma confined by
the magnetic field:
\begin{equation}
 \label{eqm:density}
    \frac{dn}{dt}= -n \frac{\partial v}{\partial s},
\end{equation}
\begin{equation}
 \label{eqm:momentum}
    n m_{\rm H} \frac{dv}{dt} = - \frac{\partial p}{\partial s}
    + n m_{\rm H} g + \frac{\partial}{\partial s}
    (\mu \frac{\partial v}{\partial s}),
\end{equation}
\begin{equation}
 \label{eqm:energy}
    \frac{d\epsilon}{dt} +(p+\epsilon) \frac{\partial v}{\partial s} =
    E_{\rm H} - n^2 \beta P(T) + \mu (\frac{\partial v}{\partial s})^2
    + \frac{\partial}{\partial s}
    (\kappa T^{5/2} \frac{\partial T}{\partial s}),
\end{equation}
with $p$ and $\epsilon$ defined by:
\begin{equation}
 \label{eqm:pressener}
    p = (1+\beta) n K_{\rm B} T \hspace{1cm}
            \epsilon =\frac{3}{2} p + n \beta \chi,
\end{equation}
where $n$ is the hydrogen number density; $s$ is the spatial coordinate
along the loop; $v$ is the plasma velocity; $m_{\rm H}$ is the mass of
hydrogen atom; $\mu$ is the effective plasma viscosity; $P(T)$ 
represents the radiative losses function per unit emission measure;
$\beta$ is the fractional ionization, i.e.\ $n_{\rm e}/n_{\rm H}$; 
$\kappa$ is the Spitzer conductivity \citep{Spitzer62}; $K_{\rm B}$ is 
the Boltzmann constant; and $\chi$ is the hydrogen ionization potential. 
$E_{\rm H}$ is an ad hoc heating function of both space and time; this 
is the main parameter we vary to study the characteristics of the 
solutions, and it will be described in detail in \S\ref{secm2a}. 
The numerical code uses an adaptive spatial grid to follow adequately 
the evolving profiles of the physical quantities, which can vary 
dramatically in the transition region.

\subsection{Initial Conditions}
As initial conditions, we consider hydrostatic loop solutions of the 
S81 model with uniform heating. In particular we have selected two 
different solutions with maximum temperatures: 
$T_{\rm max} = 3 \cdot 10^6$~K and $T_{\rm max} = 10^7$~K. 
The initial model atmosphere uses the \cite{Vernazza} model to extend 
the S81 static model to chromospheric temperatures (the minimum 
temperature is $T_{\rm min} = 4.4 \cdot 10^3$~K). 
Table~\ref{tabmod} summarizes the characteristics of the initial 
loop conditions: maximum temperature, $T_{\rm max}$; base pressure,
$p_{\rm base}$; heating per unit time and per unit volume, $E_0$; and 
the characteristic cooling time, $\tau_{\rm cool}$, according 
to \cite{Serio91}:
\begin{equation}
 \label{eq:tau}
   \tau_{\rm cool} \sim 120 \frac{L_9}{\sqrt{T_7}} .
\end{equation}
This parameter is important in many respects, e.g.\ for comparison
with the time interval between heat pulses, and other parameters of 
the simulations as discussed below.

Note that the parameters of the loop in Table~\ref{tabmod} satisfy 
the scaling laws derived from the static RTV loop model:
\begin{equation}
 \label{eq:SLpress}
    p_{\rm base} \sim \left(\frac{T_{\rm max}}{1.4 \cdot 10^3}\right)^3
                \cdot \frac{1}{L}
\end{equation}
\begin{equation}
 \label{eq:SLenergy}
    E_0 \sim 10^5 \cdot p_{\rm base}^{1.17} L^{-0.83} .
\end{equation}

Our aim is to find solutions corresponding to loops in steady state 
conditions over long time scales, as observed. 
We find that, in general, the initial conditions have little influence 
on the loop evolution driven by the repeated impulsive heating, and
that they can be important only in a region of the parameter space 
at the boundary between stability and instability. In particular, the
results can change if the loop is initially already hot and dense.

\subsection{The Heating Function} \label{secm2a}
The heating function, $E_{\rm H}(s,t)$, is assumed to be a separate 
function of space and time:
\begin{equation}
    E_{\rm H}(s,t) = H_0 \cdot g(s) \cdot f(t)
\end{equation}

\paragraph{Amount of Energy Release}---
One of our main goals is to model stellar coronae of activity levels 
from intermediate to high, showing features not reproduced by 
standard models. The characteristic plasma temperatures for such 
coronae range from a few million degrees up to $\gtrsim 10^7$~K.

The reference value for the time-averaged intensity of the heating 
is the energy required to heat the initial hydrostatic model ($E_0$ 
of Table~\ref{tabmod}). We run two sets of simulations choosing $H_0$ 
such that the heating, averaged in time and along the loop, 
$\langle E_{\rm H} \rangle$ 
($= \frac{1}{V} \frac{1}{t} \int_V \int_t E_{\rm H}(s,t) dV dt$), 
is equal to $E_0$ corresponding to $T_{\rm RTV} = 3 \times 10^6$~K 
and $T_{\rm RTV} = 10^7$~K respectively\footnote{$T_{\rm RTV}$ 
corresponds to $T_{\rm max}$ of Eq.~(\ref{eq:SLpress}). $T_{\rm RTV}$ 
and $E_0$ are linked through the scaling laws of 
Equations~(\ref{eq:SLpress})~and~(\ref{eq:SLenergy}).}. 
We also consider higher values of the heating 
($\langle E_{\rm H} \rangle = 4E_0$) because:
\begin{itemize}
\item[1.] we want to check whether unstable loops become stable with
	 stronger heating;
\item[2.] foot-point heated loops are cooler than uniformly heated
	loops with the same total energy input; we include simulations
	with stronger heating to compensate for this effect.
\end{itemize}

\paragraph{Spatial Distribution of Heating}---
The spatial distribution of the heating is described by a Gaussian
function:
\begin{equation}
 \label{eq:gauss}
    g(s) = e^{-(s-s_0)^2/(2 \sigma^2)},
\end{equation}
centered at the loop foot-points (i.e.\ $s_0 = 0$) for all the
simulations.

\begin{figure*}[!ht]
\centerline{\psfig{figure=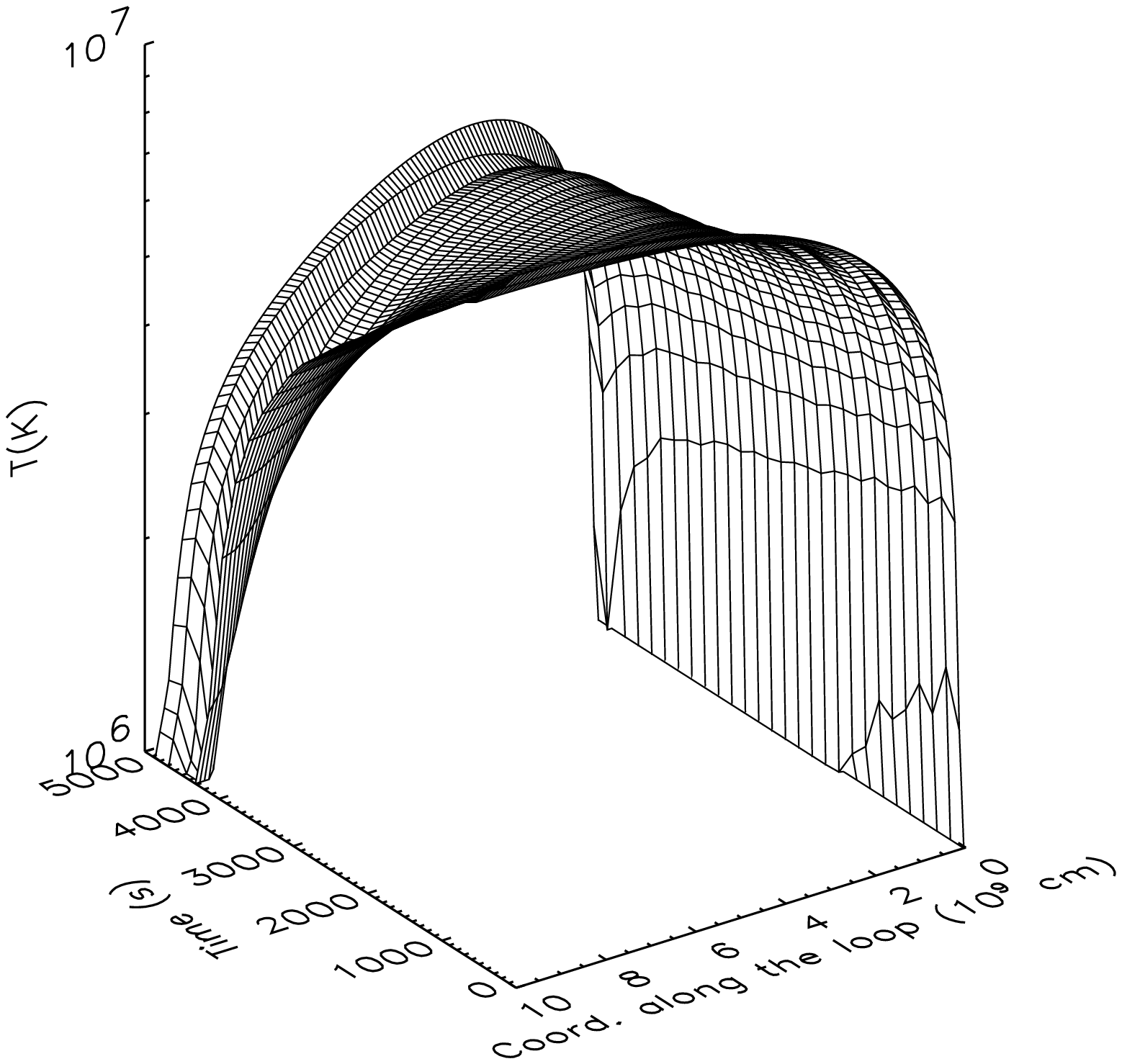,width=8cm}
	\psfig{figure=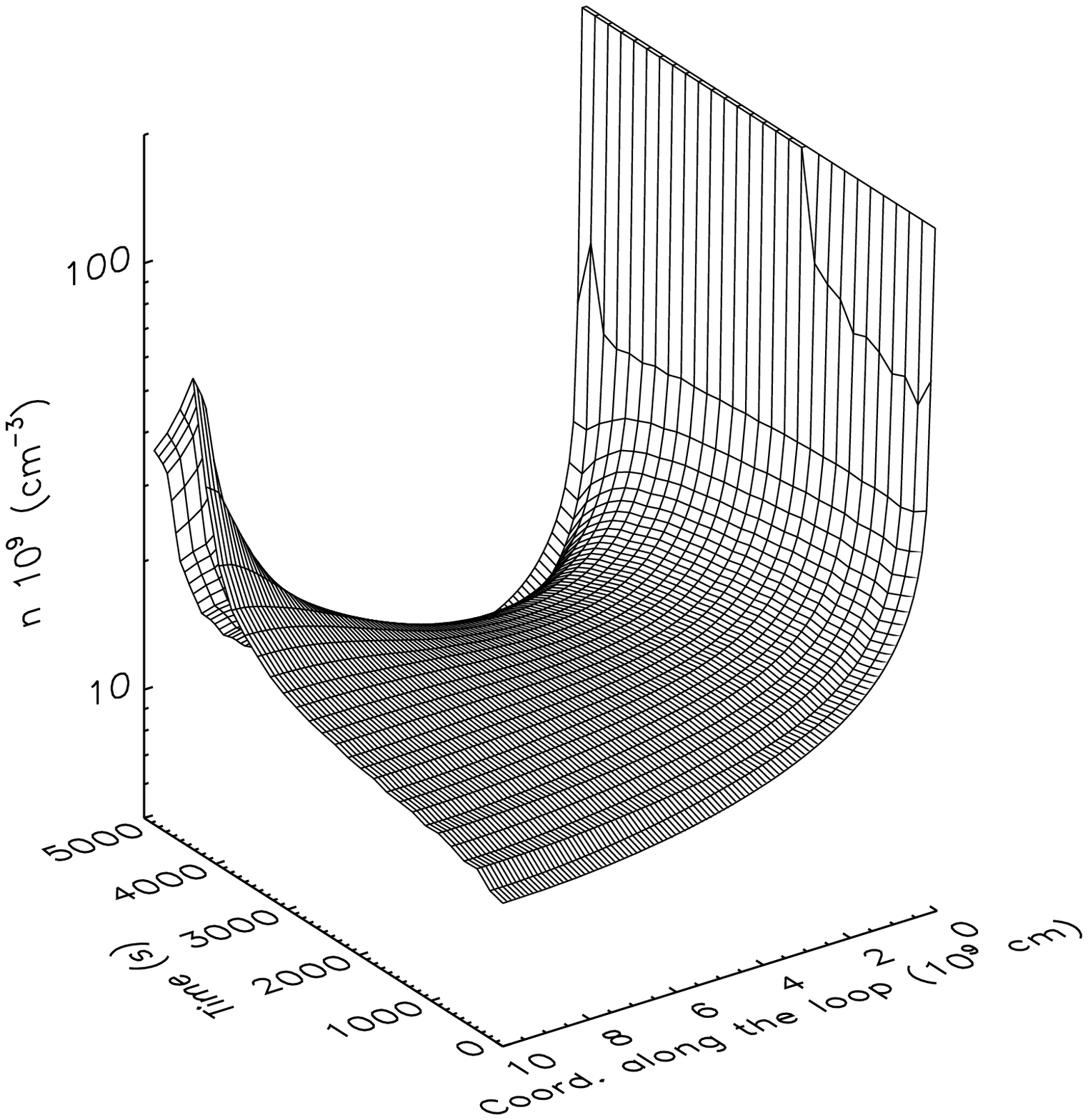,width=8cm}}
\caption{Evolution of the temperature ({\em left}) and the density 
	({\em right}) distributions along (half of) a loop with 
	semilength $L = 10^{10}$~cm, initial maximum temperature 
	$T_{\rm max}=10^7$~K, and constantly heated at its foot-points
	with $\sigma=L/3$. $T$ (in K) and $n$ (in $10^9$~cm$^{-3}$) 
	are plotted vs.\ time (in seconds) and vs.\ the 
	coordinate along the loop (in units of $10^9$~cm). 
	The loop top is on the left, and the 3D plots are 
	oriented with the initial profiles toward the observer.
	\label{fig1}}
\end{figure*}

We assume that the heat deposition is exactly the same in both
legs of the loop, thus, as mentioned above, we have symmetry about
the loop apex.

The spatial extent of the heating is a fundamental parameter of our
study, given our interest in the effect of the concentration of the 
heating on the stability and on other characteristics of the solutions. 
A preliminary set of runs has indicated that the 
boundary between stable and unstable solutions is
$\sigma \sim L/5$. Therefore, we will discuss in detail the results 
for $\sigma = L/10, L/5, L/3$, i.e.\ smaller than, equal to and larger
than the critical value. For $\sigma> L/3$ the solutions do not depart 
significantly from the static solutions.

\paragraph{Temporal Distribution of Heating Pulses}---
Given the transient nature of the simulated heating, a natural 
reference point is represented by the loop models with heating 
that produces flare events.
In these models the heating typically consists of two terms: a 
constant and uniform heating keeping the steady conditions of the 
initial loop, i.e.\ the steady coronal heating; and a transient
much larger heating which triggers the flare. 
In our analysis we neglect the constant term and assume no 
steady coronal heating. 

The heat pulses are periodic with period $\tau$, and duty cycle 10\%,
i.e.\ the heat is active only for a pulse lasting 1/10 of the period. 
Each heat pulse is a step function, constant when active and zero 
otherwise. We consider values of $\tau$ smaller than the cooling 
time of the loop, $\tau_{\rm cool}$, in order to prevent the loop 
from collapsing, but long enough to induce significant changes in 
the loop.

Our choice of periodic pulses, instead of a random distribution in 
time, allows us to limit the free parameters and to focus on the most
critical ones. We have also run test cases with random heat pulses in
order to check for possible differences \citep{Testa03}; solutions
do not differ significantly for similar pulse parameter values 
(i.e.\ average energy release and average interval between pulses). 
Also \cite{Peres93} have shown that the results do not differ 
substantially from those obtained with random pulses with 
equal average repetition time and duty cycle.

The simulations are run for at least $\sim 10$ periods and
a total time, $t_{\rm run}$, much longer than the cooling time, in 
order to be able to distinguish between stable and unstable solutions.
The total times are of the order of several hours, comparable with
the time coverage of actual coronal observations.

We summarize the parameters of the simulations in Table~\ref{tabmod}.

\section{Results} 
\label{sec3}

\subsection{Evolution of Solutions}
We know from previous works that loops continuously heated at the 
foot-points are not stable, for $\sigma < \sigma_{\rm critical}$. 
For comparison with pulse-heated loops, we first show the 
results for a loop with continuous heating.

Figure~\ref{fig1} shows the evolution of temperature and density for 
a loop evolving from initial hydrostatic conditions of 
$T_{\rm max}\sim 10^7$~K, with a foot-point heating ($\sigma = L/3$)
constant in time.
The evolution is shown as a 3D plot, where the temperature, $T$, and 
the density, $n$, are plotted vs.\ time and vs.\ the coordinate, $s$,
along the loop. $s=0$, corresponds to the loop base, $s=10^{10}$~cm, 
to the apex; the 3D plot is oriented with the initial profiles toward
the observer.
The figure shows that the plasma progressively cools down and 
condensates at the loop apex.
As discussed by e.g.\ RTV and S81, a configuration with temperature 
and density inversion ($s(T_{\rm max})$, $s(n_{\rm min})$ $< L$)
is thermally unstable.
Despite the continuous energy release, the loop collapses on time 
scales only slightly longer than the characteristic cooling time.

\begin{figure*}[!ht]
\centerline{\psfig{figure=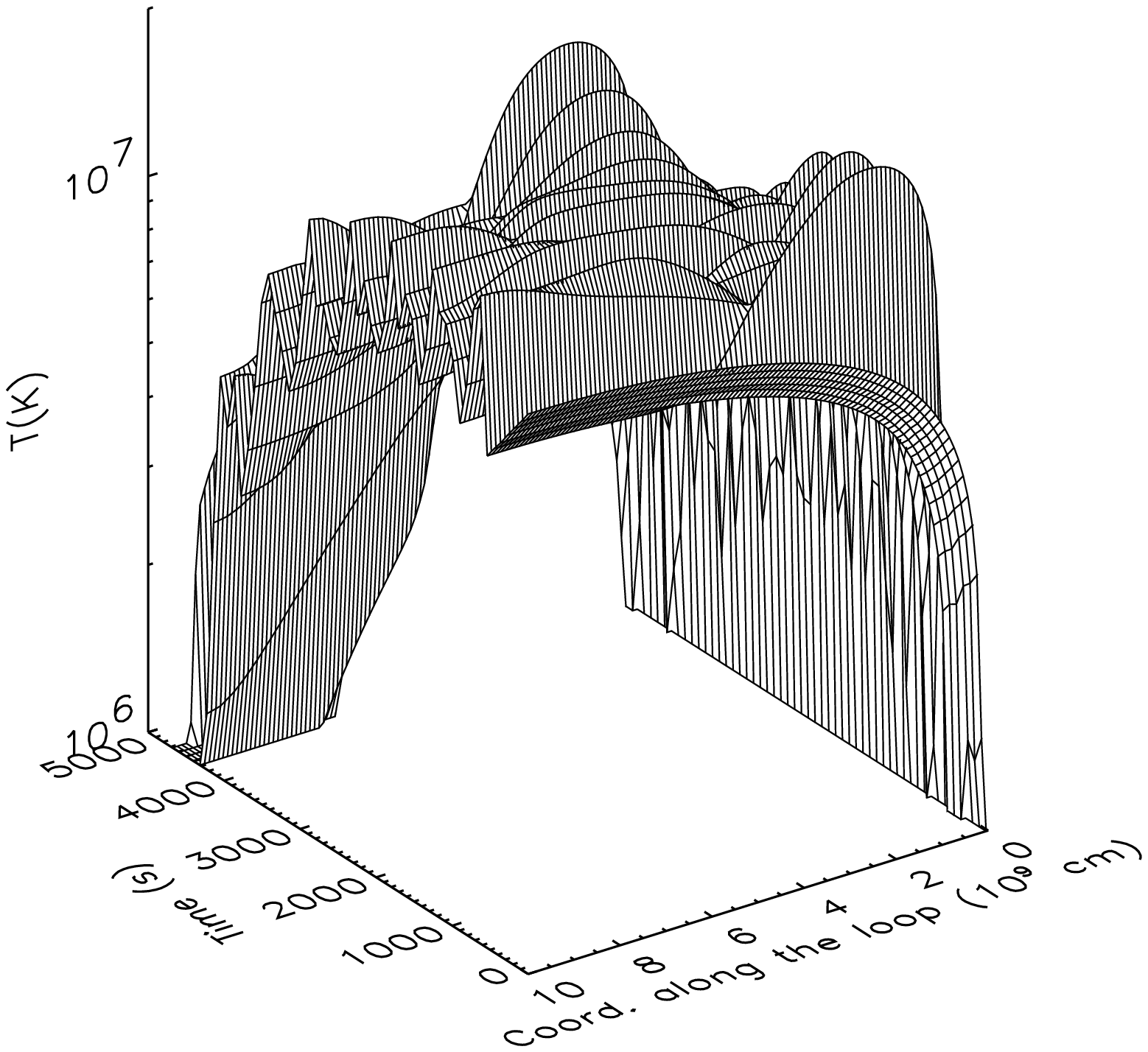,width=8cm}
	\psfig{figure=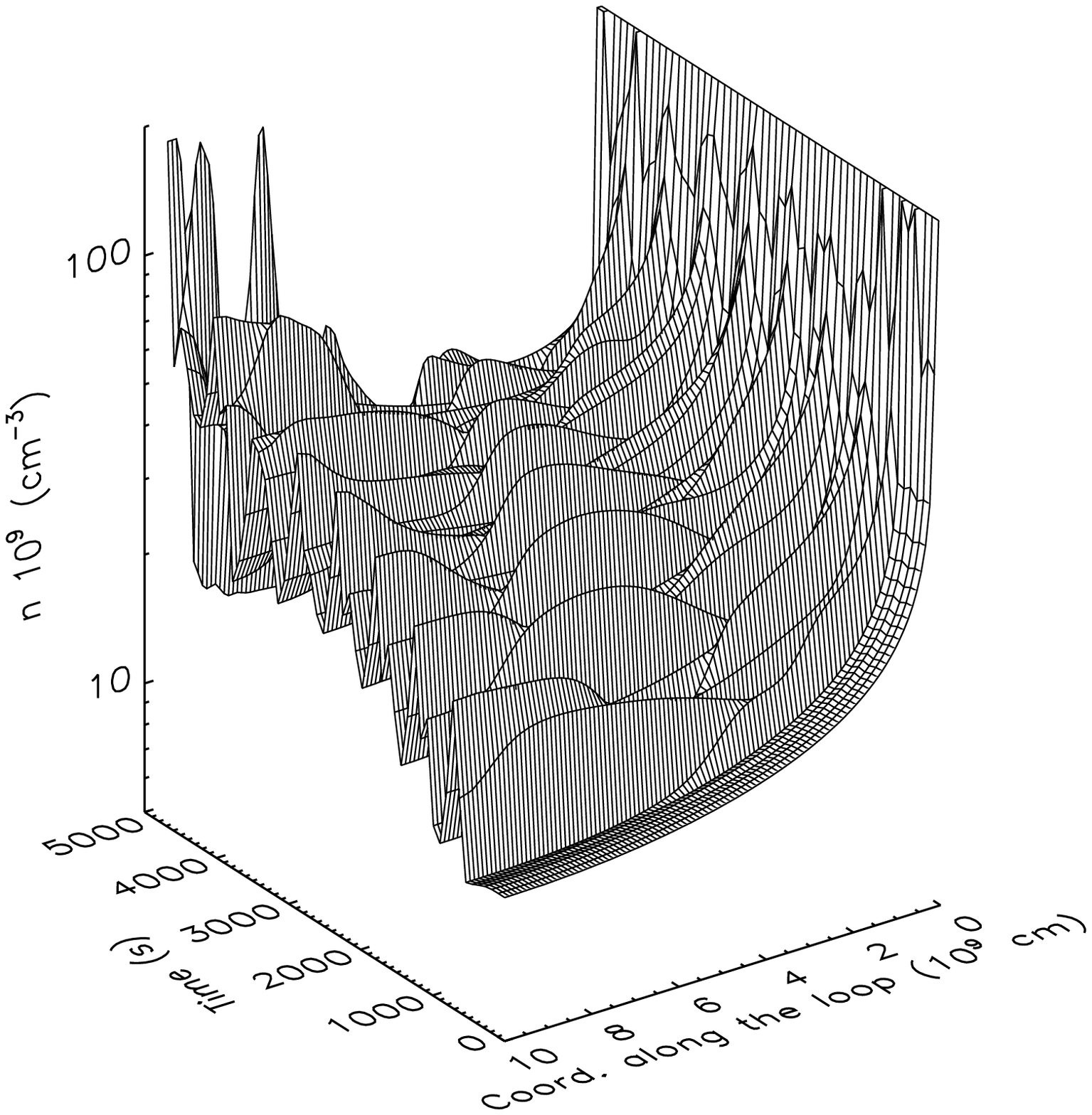,width=8cm}}
\caption{Evolution of the temperature ({\em left}) and the density 
	({\em right}) distributions along (half of) an unstable 
	pulse-heated loop.  
	The heating function parameters are: $\tau=\tau_{\rm cool}/2$, 
	$\sigma=L/10$, $\langle E_{\rm H} \rangle=4E_0$ 
	(see Table~\ref{tabmod} for their definition). 
	The 3D plots are in the same format as in Figure~\ref{fig1}.
	\label{fig2}}
\end{figure*}

\begin{figure*}[!ht]
\centerline{\psfig{figure=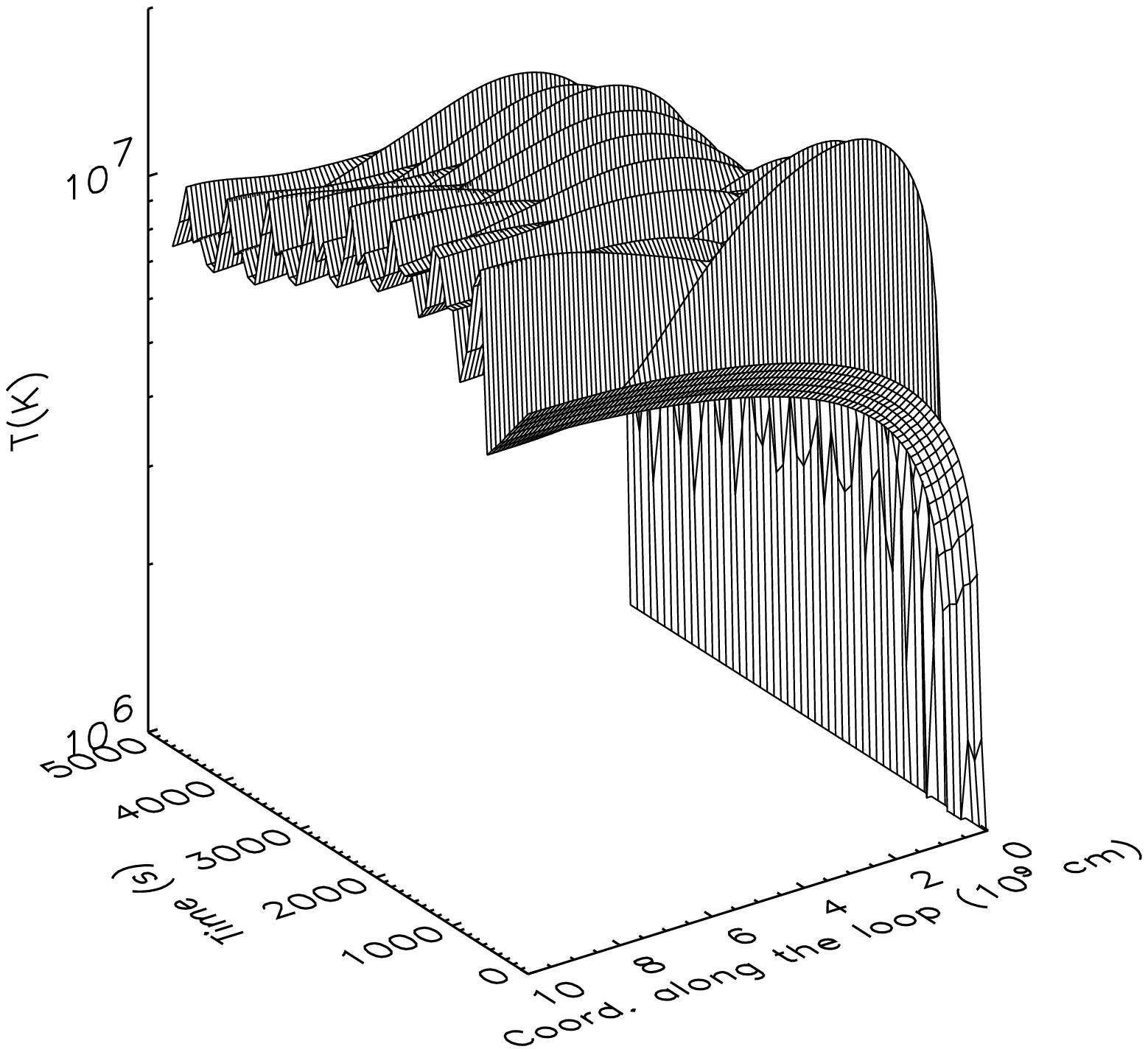,width=8cm}
	\psfig{figure=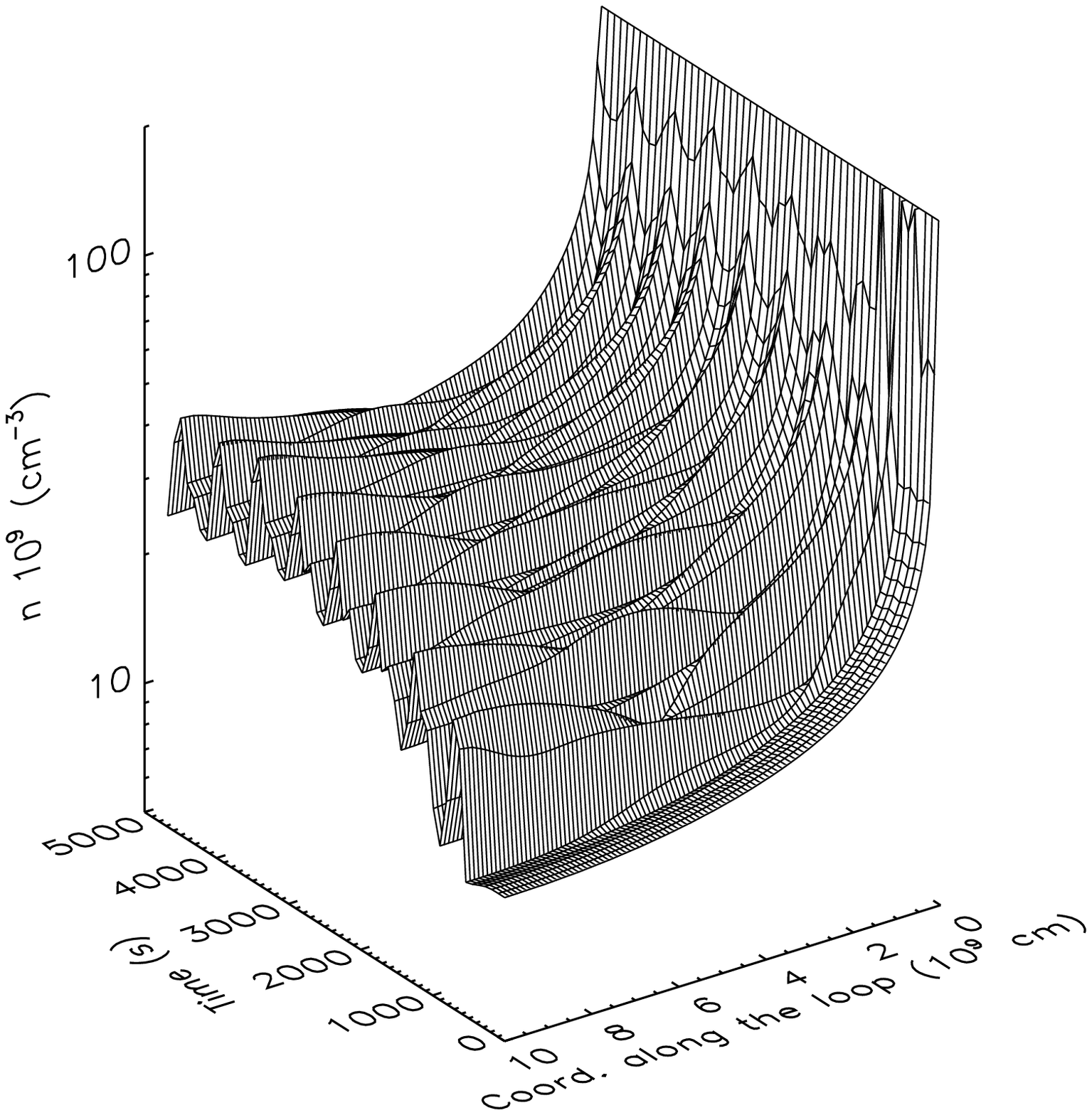,width=8cm}}
\caption{Evolution of the temperature ({\em left}) and the density 
	({\em right}) distributions for a stable pulse-heated loop. 
	The heating function parameters are: $\tau=\tau_{\rm cool}/2$, 
	$\sigma=L/5$, $\langle E_{\rm H} \rangle=4E_0$, i.e.\ the 
	parameters are the same as those of the loop of 
	Figure~\ref{fig2}, except for $\sigma$. 
	The 3D plots are in the same format as in Figure~\ref{fig1}.
	\label{fig3}}
\end{figure*}

Figure~\ref{fig2} and~\ref{fig3} show results for pulsed heating,
and in particular for an unstable and a stable loop respectively,
in the same format of Figure~\ref{fig1}. 

The solution shown in Figure~\ref{fig2} is for a loop with 
semi-length $L = 10^{10}$~cm and heating function parameters: 
$\tau=\tau_{\rm cool}/2$; $\sigma=L/10$; and 
$\langle E_{\rm H} \rangle=4 E_0$ (see \S\ref{sec2} or 
Table~\ref{tabmod} for their definition).
Initially, the solution does not depart markedly from average 
conditions, but after $t \sim 2 \tau_{\rm cool}$ the instability 
develops at the loop apex, i.e.\ the least heated place in the loop.
As the apex plasma cools slightly, the radiative losses increase. 
Since the apex plasma is not sustained by significant local heat 
deposition, it cools down even more and condenses; the increase in 
density and the decrease in temperature further enhances the 
radiative losses. 
Once the instability is triggered, the loop quickly collapses because 
the radiative losses increase for decreasing temperature and  
increasing density.

All the unstable cases show similar temperature and density
evolution; the more concentrated the heating, the faster the
instability occurs. The instability invariably starts at the apex.

Figure~\ref{fig3} shows an example of a long-lived solution;
for a more direct comparison we show the loop with the same
characteristics as that shown in Figure~\ref{fig2}, except 
that the spatial deposition of the heating is $\sigma = L/5$ instead 
of $L/10$ (Figure~\ref{fig2}).
In spite of the fluctuations due to the heat pulses, the loop 
is stable on long time scales and settles to a state of higher 
density because of the significant chromospheric evaporation
driven by the heat pulses.

\begin{figure*}[!ht]\hspace{-1cm}
\centerline{\psfig{figure=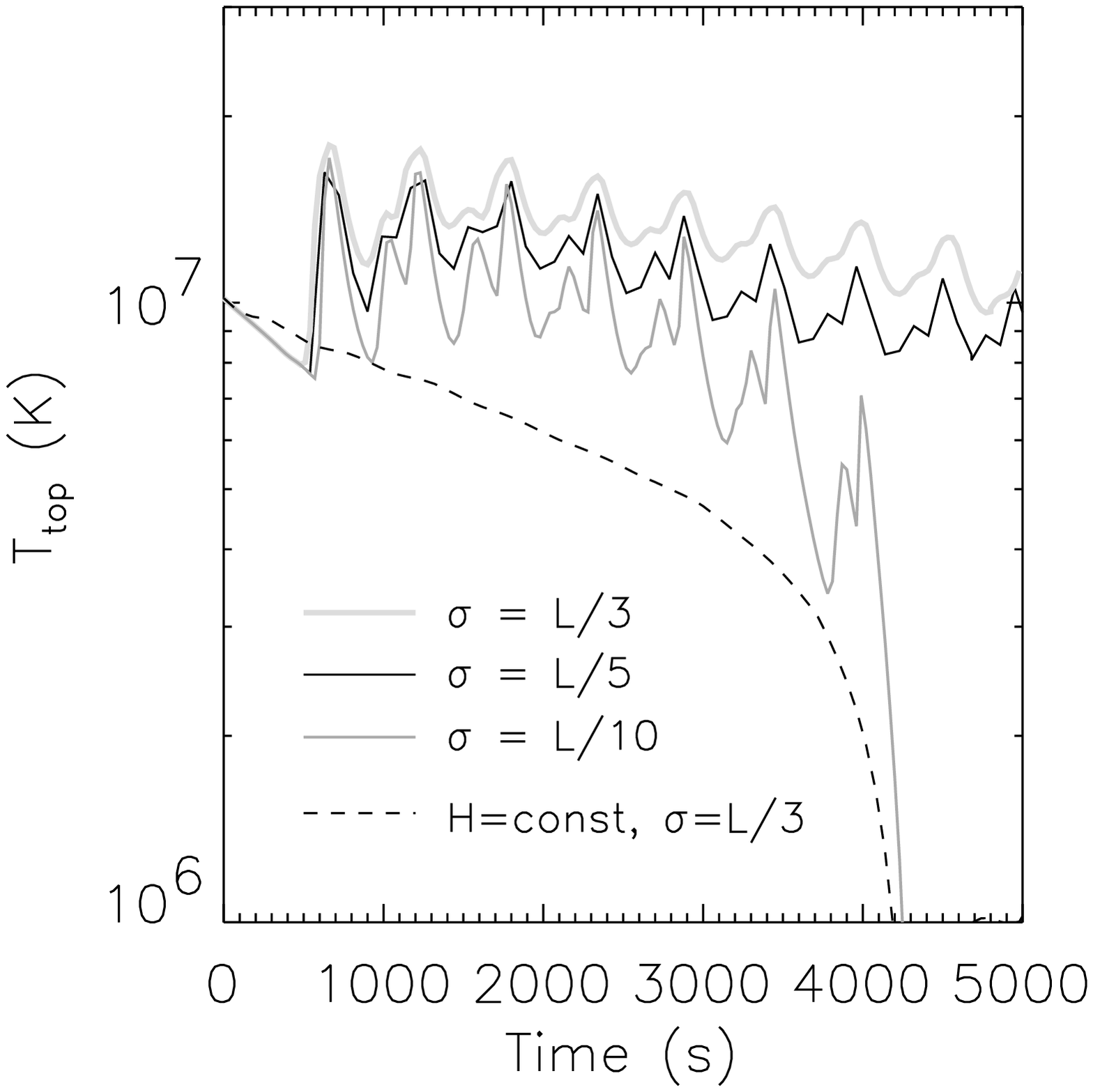,width=9cm}\hspace{-1.2cm}
       \psfig{figure=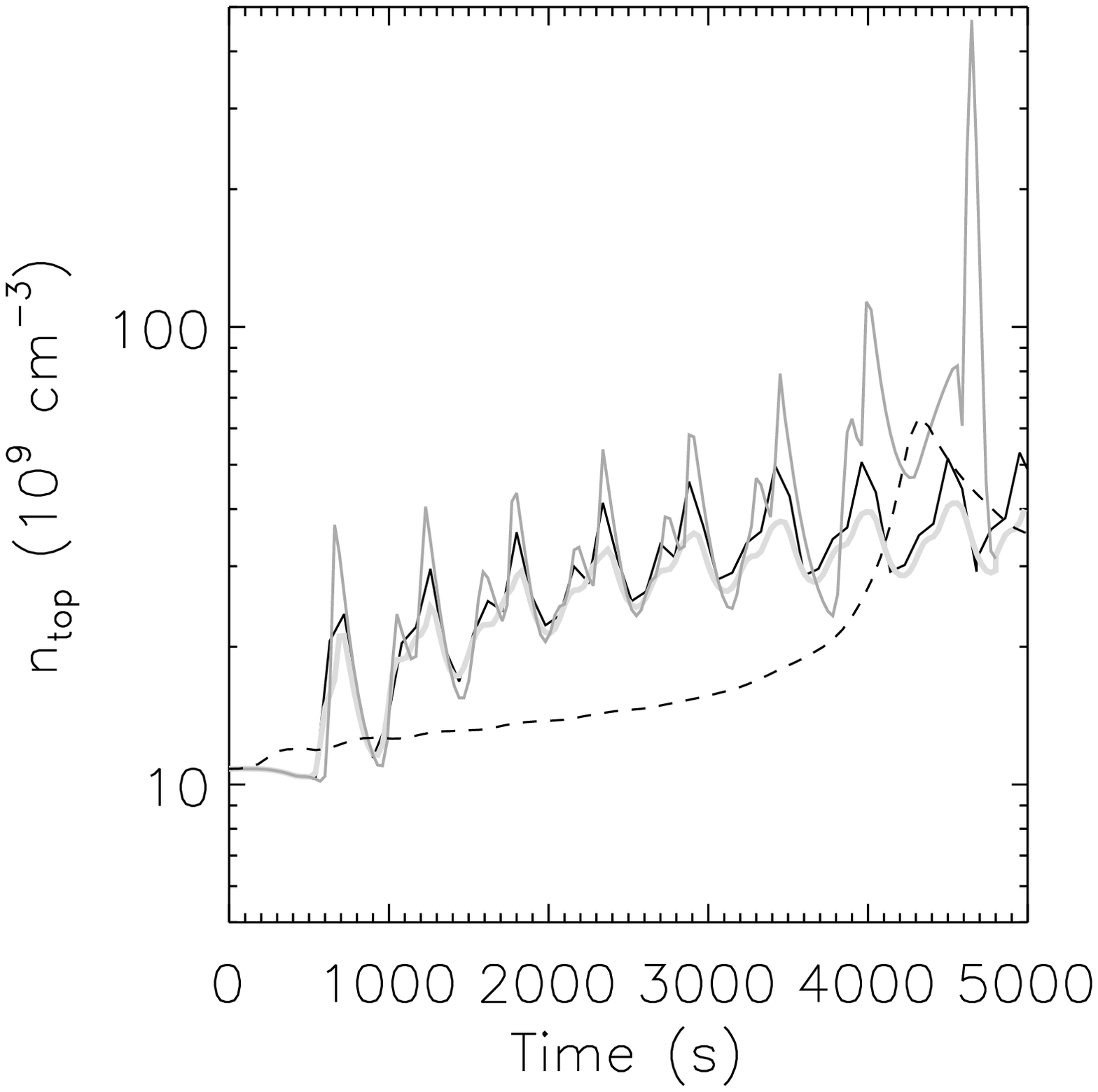,width=9cm}}\vspace{-0.7cm}
\caption{Evolution of the temperature ({\em left}) and the density 
	({\em right}) at the loop top, for different spatial and 
	temporal characteristics of heating: continuous heating 
	concentrated at the foot-point with spatial scale 
	$\sigma=L/3$ (dashed lines), periodic heat pulses with 
	$\sigma=L/3$ (thick light grey), $\sigma=L/5$ (solid black), 
	$\sigma=L/10$ (dark grey).
	\label{fig:evol}}
\end{figure*}
Figure~\ref{fig:evol} shows the evolution of the plasma temperature
and density at the loop apex for the loop models with pulsed 
heating as compared with the solution for constant heating.
The evolution of the plasma properties at the loop apex are presented 
for the loop constantly heated at the foot-point ($\sigma=L/3$;
dashed lines), and for the loops subject to heat pulses with
different spatial distribution: $\sigma=L/3$ (thick light grey); 
$\sigma=L/5$ (solid black); and $\sigma=L/10$ (dark grey).
The figure shows the instability of the constantly heated loop
and of the solution with heat too concentrated at the foot-points
($\sigma=L/10$). The two solutions with $\sigma=L/5$ and 
$\sigma=L/3$ appear more stable.

\paragraph{Effect of Heating Parameters ---}
The critical parameter for loop stability is the spatial width of the
heating function: we have stable solutions for $\sigma \gtrsim L/3$ 
and unstable solutions for $\sigma \lesssim L/10$; 
for $\sigma \sim L/5$ the loops are on the edge of stability,
and other characteristics of the heating (intensity, interval 
between pulses) become important.
We find consistent results for both the cooler and the hotter solutions.

The interval between pulses, $\tau$, does not appear to be a critical
parameter for stability as long as it is not too small 
($\tau \ll \tau_{\rm cool}$), i.e.\ too close to the constant heating
case that is unstable for $\sigma = L/3$ (see Fig.~\ref{fig1}), nor 
too long with respect to the cooling time, since the loop would then
catastrophically cool.

\subsection{Emission Measure Distribution}
Besides investigating the stability of the solutions, our modeling 
effort aims at studying the effect of shrinking the heating region at
the loop foot-points on the $EM(T)$, which is one of the main derived
quantities that allow us to study and to compare the coronal properties 
of different stars.

As described in section~\ref{sec2}, we analyze two different sets
of models with maximum temperatures of about 
$\sim 3 \times 10^6$~K, and $\sim 10^7$~K, respectively. Since we
find results that are qualitatively similar in both cases, 
we will discuss in detail the model of hotter loop, because one of
the questions we want to address is the possibility of reproducing
the $EM(T)$ of active stars (generally characterized by peak 
temperature of the order of $10^7$~K) with the proposed model.

From our simulations we can model the global emission of a stellar 
corona composed of many loops impulsively heated by microflares 
occurring close to their foot-points.
Under the assumption that all of the loops have a statistically 
analogous evolution, taking a sample of profiles at different 
times in the evolution of a single loop, as shown for example in 
Fig.~\ref{fig3}, is equivalent to observing all of the loops 
simultaneously, with each loop at a different stage in its evolution.
We take 200 outputs from each simulation, uniformly sampled in time, 
and consider each output as a snapshot of an independent loop.
Thus, the total emission from the corona, composed of all these loops, 
is the sum of the emission of all the snapshots.

From the temperature and density profiles of each output we derive
$EM(T)$ and then we sum them to obtain the global $EM(T)$.
We derive $EM(T)$ in each temperature bin, $\Delta T_i$, as 
$EM(T_i) = \sum_j n_e^2(j)~ds_j$ where $j$ spans over the spatial
bins whose temperature falls within $\Delta T_i$, and the temperature 
bins are constant in $\log T$ such that $\Delta \log T = 0.1$.
Therefore $EM(T)$ obtained with this procedure differs from the 
definition of Eq.(\ref{eq:EM}) by a normalization factor that
corresponds to the cross-sectional area of the loop.

\begin{figure}[!ht]
\centerline{\psfig{figure=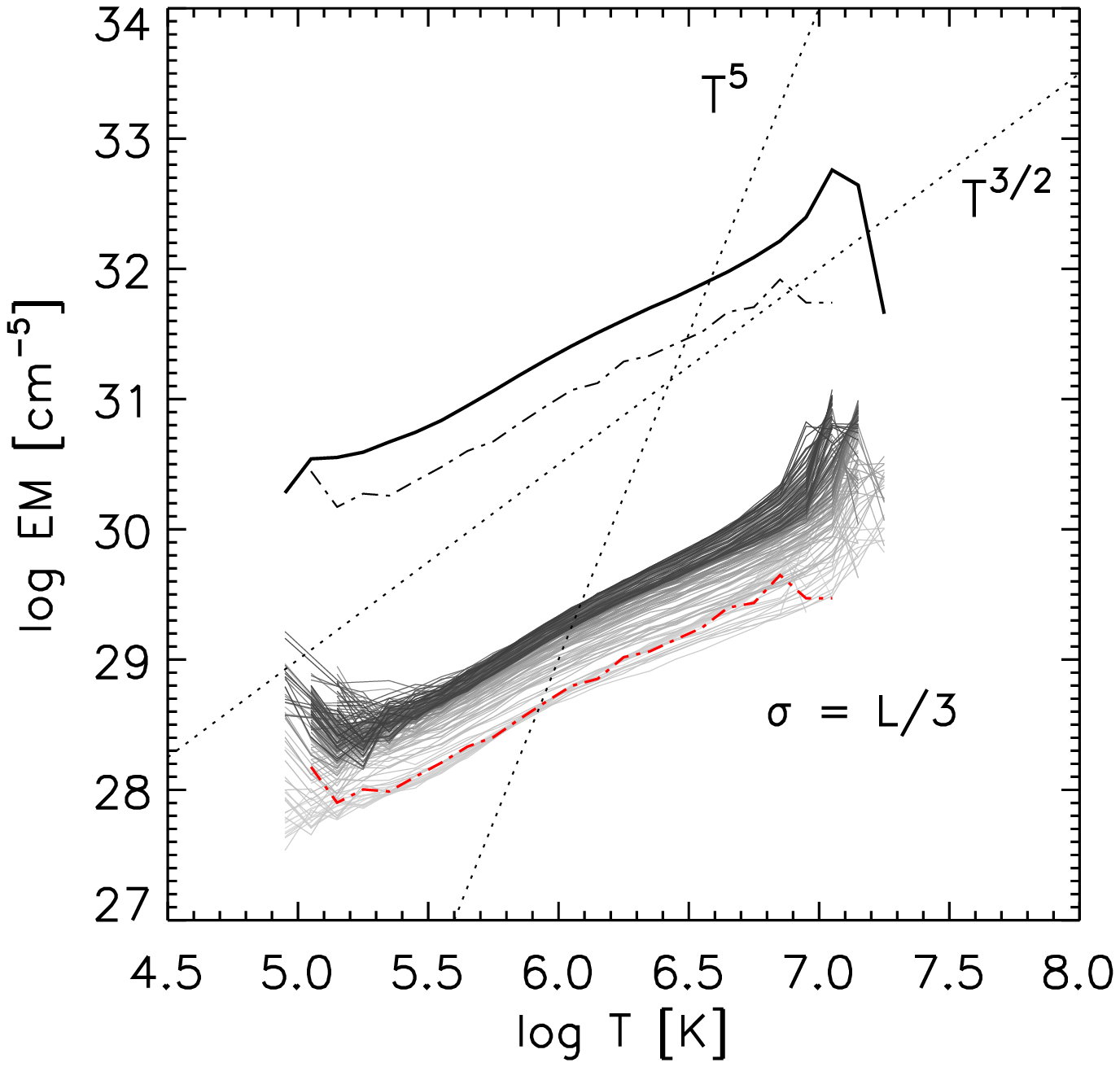,width=7cm}}\vspace{-2cm}
\centerline{\psfig{figure=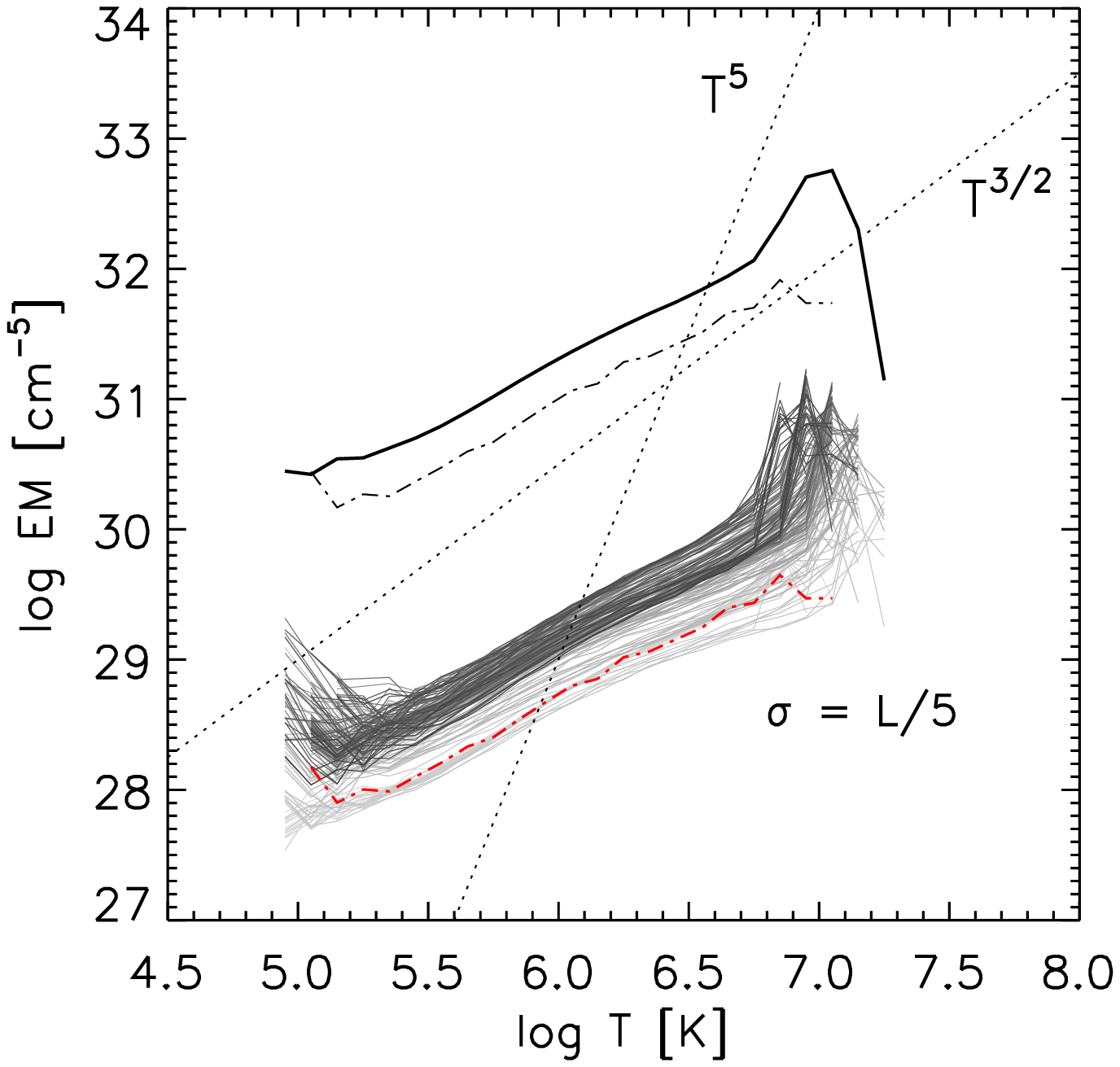,width=7cm}}\vspace{-2cm}
\centerline{\psfig{figure=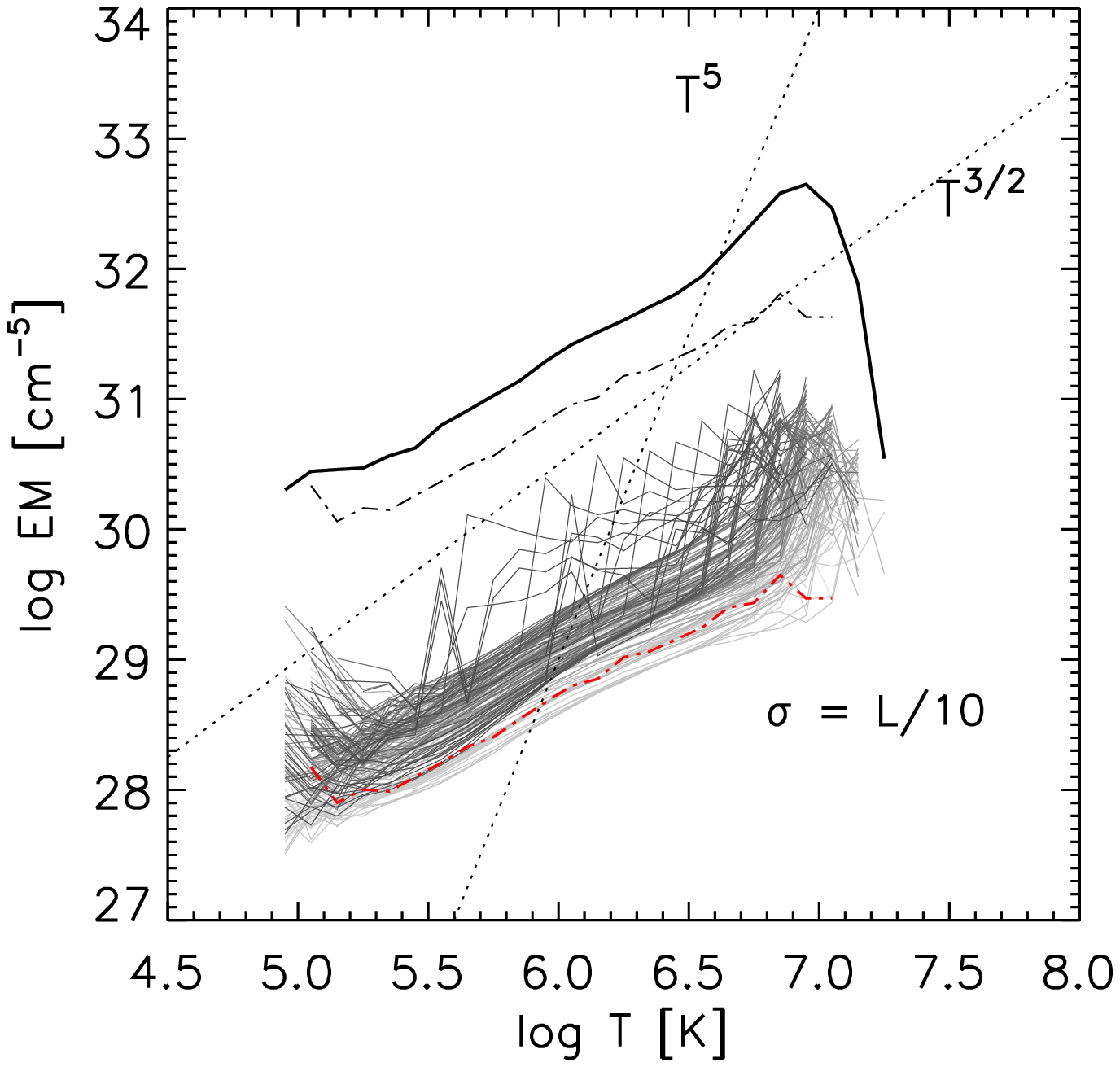,width=7cm}}
\caption{Emission measure distribution, $EM(T)$, for the solutions 
	with $T_{\rm max} \sim 10^7$K, $\langle E_{\rm H} \rangle=4E_0$, 
	and $\sigma=L/3$ ({\em top}), $\sigma=L/5$ ({\em middle}), 
	$\sigma=L/10$ ({\em bottom}). 
	The lines in the lower part of the plots correspond to the 
	$EM(T)$ of the individual outputs, regularly sampling the 
	whole loop evolution; the lines get darker from $t=0$ (lighter 
	gray) to $t=t_{\rm run}$ (darker gray lines). 
	The dashed-dotted lines correspond to the $EM(T)$ of the
	initial static condition, while the upper thick black line
	is the total $EM(T)$. For comparison the dotted lines indicate 
	the power laws corresponding to	$EM(T) \propto T^{3/2}$ and 
	$EM(T) \propto T^{5}$.
	\label{fig:emt}}
\end{figure}

Figure~\ref{fig:emt} shows $EM(T)$ for the loop models with average 
coronal temperatures $\sim 10^7$~K, and heating functions with 
different spatial widths. The solutions with $\sigma=L/3$ ({\em top}), 
$\sigma=L/5$ ({\em middle}), and $\sigma=L/10$ ({\em bottom}), show 
the effect on $EM(T)$ of varying the heating concentration.
The lines in the lower part of the plots of Figure~\ref{fig:emt} 
correspond to $EM(T)$ for each output represented with
lines darkening from $t=0$ (light gray) to $t=t_{\rm run}$ 
(dark gray). The upper thick black line represents the total $EM(T)$, 
which is the sum of all the individual distributions.
Therefore the total $EM(T)$ is the one that 
would characterize a corona composed entirely of such loops.
In order to allow for an easier comparison of the static model with 
the dynamic simulations, $EM(T)$ of the initial static solution 
(dashed-dotted line) is also shown arbitrarily shifted next to 
the total $EM(T)$.  
In Figure~\ref{fig:emt} the power laws corresponding to 
$EM(T) \propto T^{3/2}$ and $EM(T) \propto T^{5}$ are also plotted 
as useful points of reference, as discussed in \S\ref{intro}.

The $EM(T)$ of the initial hydrostatic solution is quite flat, close 
to the $T^{3/2}$ power-law.  Figure~\ref{fig:emt} shows that $EM(T)$ 
can clearly be used to characterize the dynamic simulations since we 
see that the total $EM(T)$ changes consistently as the heating becomes 
more concentrated in the foot-point region.
The main difference of the total $EM(T)$ (solid lines) from the 
standard hydrostatic distribution is the presence of a well-defined 
peak at high temperatures ($\sim 10^7$~K).
This is also the temperature range corresponding to the bulk of the 
emission from active stellar coronae. 

For $\sigma = L/3$ the $EM(T)$ is close to the initial curve. 
Narrowing the region of energy release to $\sigma = L/5$ causes the
$EM(T)$ to steepen and approach the scaling derived from stellar 
observations. The more concentrated the heating, the wider the peak. 

The $EM(T)$ does not depend on the parameter $\tau$ as it does on 
the concentration of the heating.

\begin{figure*}[!ht]
\centerline{\psfig{figure=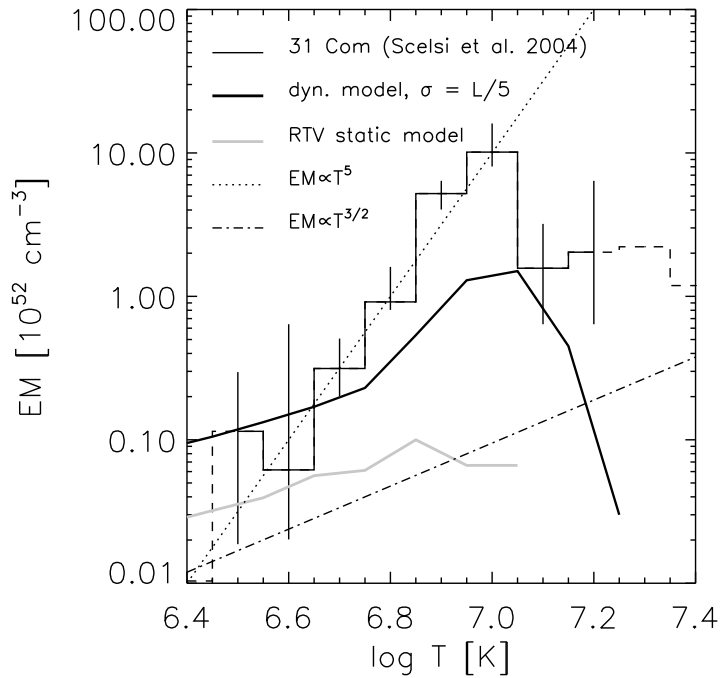,width=10cm}}
\caption{$EM(T)$ derived for a loop model with maximum temperature
	$\sim 10^7$~K, and foot-point heating ($\sigma = L/5$;
	thick line), as compared to the $EM(T)$	of 31~Com (histogram) 
	derived from an X-ray spectrum. 
	As a reference, the $EM(T)$ for an RTV hydrostatic model 
	(light grey line), and power-laws with $EM(T) \propto T^5$ 
	(dotted line) and $EM(T) \propto T^{3/2}$ (dash-dotted line) 
	are shown as well. $EM(T)$ from models are arbitrarily scaled.
	\label{fig:obsvsm}}
\end{figure*}

In Figure~\ref{fig:obsvsm} we compare the $EM(T)$ derived from our 
model, for $\sigma = L/5$, to an $EM(T)$ derived by \cite{Scelsi04} 
from {\em XMM-Newton} spectral observations of the active star 31~Com.
The $EM(T)$ derived from a RTV hydrostatic model is also shown.
The $EM(T)$ derived from our model, with impulsive foot-point heating, 
is qualitatively similar to the $EM(T)$ of 31~Com.

\section{Discussion and Conclusions}
\label{sec4}

Coronal loops are known to be unstable if heated continuously in a 
sufficiently small region at their foot-points. Also, the $EM(T)$ 
of stable, uniformly-heated loops of constant cross-section 
is known to obey a power law with index $\sim 3/2$. 
In the present work, we have investigated the existence of long-lived
coronal loops heated by sequences of pulses located at their 
foot-points and studied the changes induced in the $EM(T)$ by this
form of heating. 
In particular, we examined relatively long ($2 \times 10^{10}$~cm) 
and hot ($\approx 3 \times 10^6$ K and $10^7$ K) loops appropriate 
to investigating both the solar environment, which we can spatially 
resolve and compare directly with models, and the coronae of active 
stars, where evidence of different $EM(T)$ has been found.

We find long-lived dynamically stable solutions in cases with heating
concentrated close to the foot-points, with values of $\sigma$ down 
to $\sim L/5$. Therefore we find stable solutions for much more 
concentrated heating compared with the solutions of the S81 model; 
the latter, using a heating spatial distribution of the form
$H_0 e^{-s/s_{\rm H}}$, does not yield stable solutions for 
$s_{\rm H} \sim L/3$.
The value of $\sigma \sim L/5$ represents the border zone between
rapidly unstable and long-lived loops for both sets of simulations 
with different maximum temperatures (see Tab.~\ref{tabmod}), provided
that the heating is strong enough to sustain the loop.
We note, however, that close to this boundary some unstable cases
appear quasi-steady when observed over time scales that are long  
compared to the cooling time.
We find that the time scales of the loops' evolution do not 
critically depend on the intensity of the heating, or on the period 
of heat pulses, with the caveats discussed in the previous section.

Our work has therefore proven that long-lived footpoint-heated loops 
can exist, provided that the heating is intermittent with the 
appropriate period -- a fraction of the loop cooling time -- which 
should be long enough to depart from continuous heating, but shorter 
than the loop plasma cooling time.  
The intermittent heating allows the plasma at the loop apex to drain
downward to the chromospheric region (as confirmed by the downward 
velocity of the plasma close to the apex in between the pulses), and
prevents the accumulation of the plasma at the loop top, and thus the 
thermal instability.
Our results are in agreement with those of other works modeling 
footpoint-heated loops \cite[e.g.,][]{Mueller04}.

In the present work we also showed that the $EM(T)$ of foot-point 
heated loops are significantly different to those of conventional 
hydrostatic loops. The $EM(T)$ of foot-point heated loops shows a 
well-defined peak, which becomes wider and wider as the heating 
becomes more concentrated towards the foot-points. 
This holds true for both sets of simulations at different temperatures 
analyzed here (see Tab.~\ref{tabmod}). 
The low temperature side of the peaks has a steeper slope than the 
$EM(T) \propto T^{3/2}$ of the static case.  The slope found for 
the stable cases with smaller $\sigma$ approaches the power law of 
$EM(T) \propto T^5$.  
We note, in passing, that loops with coronal cross-section larger
than in their chromospheres may give similar results 
\citep[e.g.,][]{Schrijver89,Ciaravella96,Sim03}. 
We plan to investigate this possibility in the future. 

The $EM(T)$ of the impulsively heated loops modeled here show a 
qualitative agreement with those derived from several X-ray and EUV
spectroscopic observations of active stars.  Recent analyses outline
a scenario in which hotter ($T \sim 10^7$~K) plasma is characterized 
by physical conditions that are fundamentally different from the 
cooler ($T \lesssim 3 \times 10^6$~K) plasma, which probably belongs 
to a different class of structures. In particular, there is increasing 
observational evidence of hot plasma with density two orders of 
magnitude higher than the cooler plasma. These density and filling 
factor results \citep[e.g.,][]{Testa04b} support a scenario in which, 
for increasing activity levels, significant flaring activity may be 
present, yielding the hotter plasma structures.  In such a scenario, 
it is likely that hot loops are sustained by impulsive energy release.  
Therefore, the theoretical model discussed here is based on 
assumptions that are consistent with the observational evidence, 
while reproducing the steep and peaked $EM(T)$ widely found for 
active stars.

There is room for several improvements to be made to our model:
we plan to model loops with a larger cross-section in the corona;
and we will consider the larger set of models for unstable, 
foot-point heated loops, since the superposition of several loops of
this kind may help to explain unresolved, monolithic loop structures
composed of many strands, or even whole stellar coronae.

\begin{acknowledgements}
We thank an anonymous referee for accurate and extensive revision 
of the paper and for many suggestions. The authors acknowledge 
support for this work from Agenzia Spaziale Italiana and Ministero 
dell'Istruzione, Universit\`a e Ricerca.
\end{acknowledgements}

\end{document}